\documentclass[aps,prb,preprint,superscriptaddress]{revtex4-2}

\usepackage{amssymb}
\usepackage{upgreek}
\usepackage{color}
\usepackage{graphicx}
\usepackage{amsmath}
\usepackage{hyperref}
\hypersetup{hypertex=true,
	colorlinks=true,
	linkcolor=blue,
	anchorcolor=blue,
	citecolor=blue}

\begin{document}

%\preprint{}

\title{Flexural wave energy harvesting by the topological interface state of a phononic crystal beam}


\author{Tian-Xue Ma}
\affiliation{Department of Civil Engineering, University of Siegen, Siegen, D-57076, Germany}
\affiliation{Institute of Engineering Mechanics, Beijing Jiaotong University, Beijing 100044, PR China}

\author{Quan-Shui Fan}
\affiliation{Institute of Engineering Mechanics, Beijing Jiaotong University, Beijing 100044, PR China}

\author{Chuanzeng Zhang}
\email[Corresponding author: ]{c.zhang@uni-siegen.de}
\affiliation{Department of Civil Engineering, University of Siegen, Siegen, D-57076, Germany}

\author{Yue-Sheng Wang}
\email[Corresponding author: ]{yswang@tju.edu.cn}
\affiliation{Institute of Engineering Mechanics, Beijing Jiaotong University, Beijing 100044, PR China}
\affiliation{School of Mechanical Engineering, Tianjin University, Tianjin 300350, PR China}


%\date{\today}

\begin{abstract}
In this study, we design the 3D-printed phononic crystal (PnC) beam with the topological interface state for harvesting the mechanical energy of flexural waves. The PnC beam is formed by arranging periodic grooves on its surface. The PnC beam with either topologically trivial or non-trivial phase can be achieved via changing the distance between the grooves. The topological interface state is then generated by combining two PnCs with distinct topological phases. The existence of the interface state of the PnC beam is verified both numerically and experimentally. To convert the mechanical energy into the electricity, a piezoelectric disc is attached at the interface of the proposed PnC beam. Compared to the reference beam harvester, the measured output power is significantly amplified by the PnC harvester at the frequency corresponding to the interface state. Furthermore, the PnC beam energy harvester based on the topological state exhibits robustness against geometrical disorders. 
\end{abstract}


\maketitle

\section{Introduction}
\label{sec1}

Topological insulators provide a unique and robust platform to achieve novel devices which are immune to back-scattering and fabrication imperfection. This concept was originally proposed in quantum systems \cite{hasan2010colloquium,qi2011topological}, and later on has been extended to several classical wave systems, including electromagnetic \cite{lu2014topological,ozawa2019topological}, acoustic \cite{xiao2015geometric,lu2017observation,zhang2017topological} and elastic \cite{cha2018experimental,chen2019mechanical,fan2019elastic} waves.

In the regime of elastic waves, the topological insulators can be realized by using artificial periodic materials, i.e., phononic crystals (PnCs) or metamaterials (MMs). The PnCs and MMs exhibit unprecedented properties which cannot be found in natural materials, such as band-gaps \cite{laude2015phononic}, negative refraction \cite{zhu2014negative} and self-collimation \cite{liu2010collimation}, to name a few. In two-dimensional (2D) PnCs, a topologically protected interface/edge state emerges at the interface between two crystals with different topological properties (topologically non-trivial and trivial). Unlike conventional guided wave modes, the topological interface states feature robustness, low loss, and unidirectionality. Generally, three methods can be used to construct 2D elastic topological insulators, including the analogs to the quantum Hall effect \cite{wang2015topological,chen2019mechanicalPRAppl}, the quantum spin Hall effect \cite{mousavi2015topologically,yu2018elastic,miniaci2018experimental,huo2018topologically}, and the quantum valley Hall effect \cite{yan2018chip,jin2020topological,gao2021broadband}. Additionally, in one-dimensional (1D) PnCs, a combination of the 1D PnCs with distinct topological characteristics supports the topological interface state, which is highly confined near the interface and robust against local disorders or defects \cite{kim2017topologically,chen2018study,yin2018band,zhou2020actively,zhou2020activelyIJMS,jin2020asymmetric,zheng2021multiple}. Thus, the interface state of the 1D PnCs exhibits potential applications in energy harvesting, sensing, and wave filtering.

One of the important applications based on PnCs is piezoelectric energy harvesting, that converts mechanical energy into electrical energy. The harvested energy can be employed to power small electronic devices, e.g., wireless sensors. One strategy for designing energy harvesters is introducing point defects into a perfect PnC \cite{carrara2013metamaterial,qi2016acoustic,ma2020flexural,lee2020enhanced,jo2020elastic,jo2020graded}. The elastic waves can be confined within the defects and hence the wave energy can be harvested by using piezoelectric materials. Another strategy is designing gradient index lenses \cite{yi2017smart,tol20193d,hyun2019gradient} or Luneburg lenses \cite{tol2017phononic} by spatially varying the effective refractive index of the PnC unit-cells. Additionally, several researches have been devoted to
design elliptical and parabolic mirrors for energy harvesting \cite{carrara2013metamaterial,carrara2012dramatic}. It is pointed out that the introduction of the topological insulators can provide new possibilities in designing PnC energy harvesters. Fan et al. \cite{fan2019acoustic} reported the harvesting of acoustic energy based on the topological interface state of a 1D PnC tube. Lan et al. \cite{lan2021energy} exploited the topological interface state for low-frequency vibration energy harvesting. Very recently, Wen et al. \cite{wen2022topological} proposed to use topological cavities in phononic thin plates for robust energy harvesting. However, the above-mentioned studies are exclusively numerical or theoretical works, while the experimental validations of piezoelectric energy harvesters based on the topological state are rarely reported.

In this study, the 3D-printed PnC beam with the topological interface state for flexural waves is firstly proposed. Then the experiments are carried out to prove the feasibility of the topological interface state for piezoelectric energy harvesting.  

\section{Model and analysis}
\label{sec2}

\begin{figure*}[ht]
	\centering
	\includegraphics{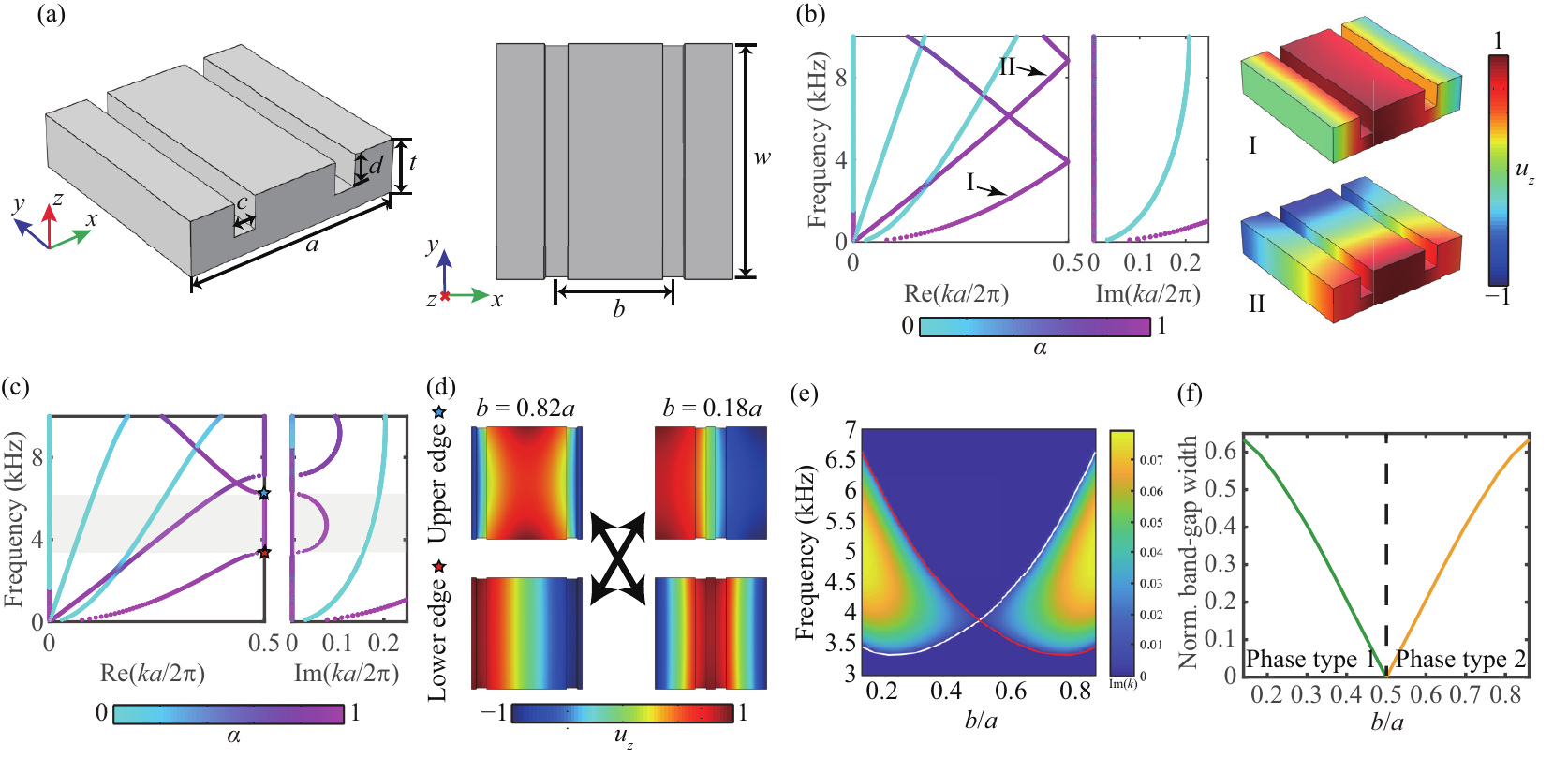}
	\caption{\label{Figure1}(a) Schematic of the unit-cell of the PnC beam and its 2D view. (b) Complex band structure of the PnC unit-cell with the geometrical parameter $b = 0.5a$, and the displacement field ($u_z$) distributions of the flexural modes at the wave number of $ka/2\pi = 0.45$. (c) Complex band structure of the PnC unit-cell with the geometrical parameter $b = 0.82a$ (or $0.18a$), where the band-gap for the symmetrically flexural mode is indicated by the shaded area. (d) Displacement field ($u_z$) distributions of the band-gap edge states at the Brillouin zone boundary ($ka/2\pi = 0.5$). (e) Variation of the band-gap edges for the symmetrically flexural mode as a function of $b/a$, where the solid lines denote the band-gap edges, and the color represents the value of the imaginary part of the wave number. (f) Variation of the normalized band-gap width for the symmetrically flexural mode as a function of $b/a$.}
\end{figure*}

The schematic diagram of the PnC beam is shown in Fig. \ref{Figure1}(a). Periodic grooves are formed on an elastic beam with the width and thickness represented by $w$ and $t$, respectively. The other geometrical parameters of the PnC beam include the lattice constant $a$, the length and depth of the grooves $c$ and $d$, and the distance between the centers of the grooves $b$.

In the present work, the weak-form of the elastodynamic equations is derived and then discretized by the finite element method (FEM) using COMSOL Multiphysics. The details are given in the supplementary material. The PnC beam is fabricated by the 3D printing with the photopolymer material, whose material parameters are as follows: the mass density $\rho = 1200$ $\rm{kg/m}^3$, the Poisson’s ratio $\nu = 0.41$, and the Young's modulus $E = 3.3$ GPa).

The complex band structure of the PnC beam is shown in Fig. \ref{Figure1}(b), where the geometrical parameters are $a = 20$ mm, $w = 20$ mm, $t = 5$ mm, $c = 2$ mm, $d = 3$ mm and $b = 10$ mm. In this work, we aim to harvest the energy of flexural waves. Therefore, to distinguish the flexural mode from the other wave modes, the polarization ratio $\alpha$ is defined as \cite{ma2020flexural}
\begin{equation}
	\label{eq1}
	\alpha  = \frac{{\int_V {{u_z}u_z^*dV} }}{{\int_V {\left( {{u_x}u_x^* + {u_y}u_y^* + {u_z}u_z^*} \right)dV} }},
\end{equation}
where $u_i \left( i=x,y,z\right)$ denotes the component of the displacement vector; the asterisk * represents the complex conjugate; and $V$ denotes the whole domain of the PnC unit-cell. According to the polarization ratio, two bands for the flexural waves (namely I and II, respectively) appear in the band structure. In the case of $b = 0.5a$ (10 mm), the unit-cell (Fig \ref{Figure1}(a)) is twice of the primitive cell (i.e., the half of the unit-cell which contains one groove in size). As a result, the band folding occurs at the Brillouin zone boundary ($ka/2\pi = 0.5$). It is pointed out that the degeneracy at the band crossing point ($ka/2\pi = 0.5$) can be lifted by changing the geometrical parameter $b$ and then a band-gap will emerge.

The displacement field ($u_z$) distributions of bands I and II are also plotted in Fig. \ref{Figure1}(b). It is observed that band I exhibits a symmetric mode with respect to the mirror plane (i.e., the $y = 0$ plane). On the other hand, the displacement field distribution of band II is anti-symmetric with respect to the mirror plane. In practice, compared with the symmetric one the anti-symmetric mode is quite difficult to excite. Thus, the symmetrically flexural mode will be considered in the following discussions.

The complex band structure of the PnC beam with $b = 0.82a$ or $0.18a$ is shown in Fig. \ref{Figure1}(c). It is clearly seen that as the distance between the grooves changes, the degeneracy at the Brillouin zone boundary separates and thus the band-gap for the symmetrically flexural mode appears. This band-gap ranges from 3.36 to 6.20 kHz with the normalized band-gap width of 0.594 (i.e., the ratio between the band-gap width and the mid-gap frequency). Further, from the imaginary part of the wave number it can be inferred that the opened band-gap is caused by the Bragg scattering of the flexural waves.

\begin{figure*}[ht]
	\centering
	\includegraphics{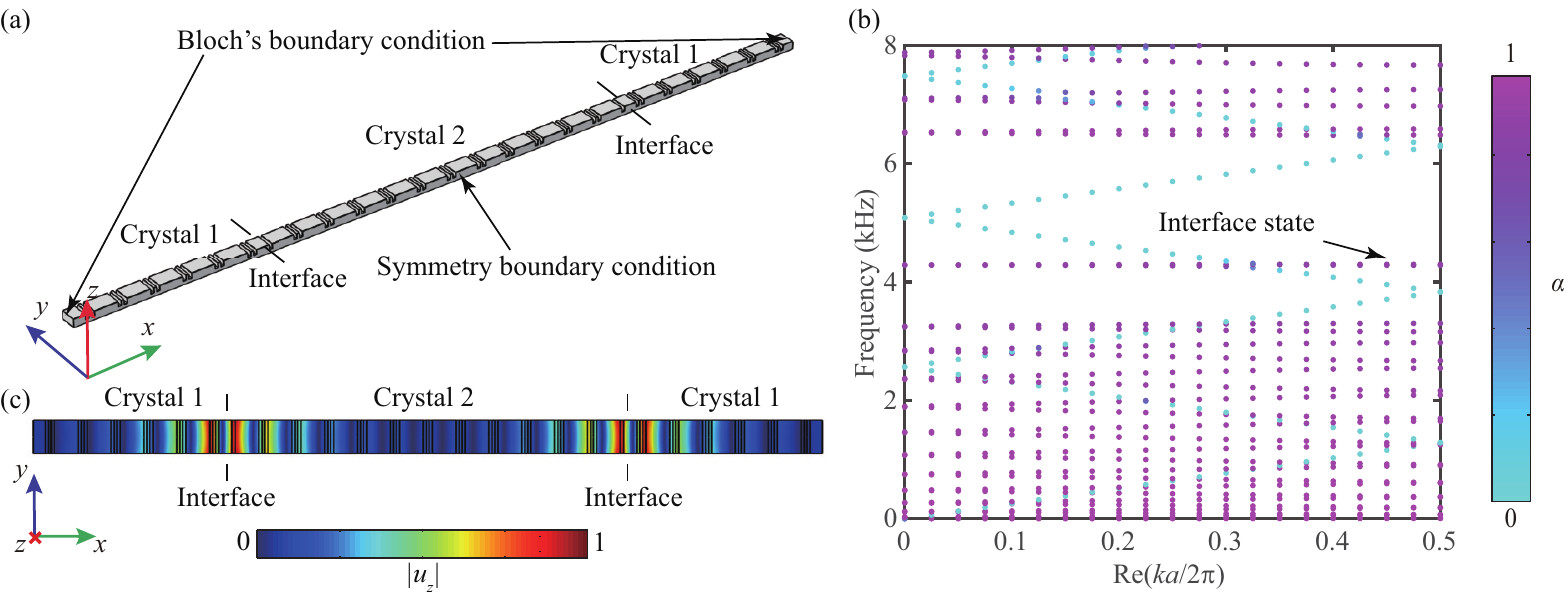}
	\caption{\label{Figure2}(a) Schematic of a PnC ribbon consisting of crystal 2 ($b = 0.82a$) sandwiched by crystal 1 ($b = 0.18a$) on its left and right sides. (b) Band structure of the PnC ribbon, where only the symmetric modes are plotted due to the symmetric boundary condition. (c) Displacement field ($|u_z|$) distribution of the topological interface state at the wave number of $ka/2\pi = 0$.}
\end{figure*}

Notice that the band structures for $b = 0.82a$ and $b = 0.18a$ are identical. However, the topological phases of the two configurations are distinct. The displacement field ($u_z$) distributions of the PnC beam at the band-gap edges ($ka/2\pi = 0.5$) are plotted in Fig. \ref{Figure1}(d). When $b$ is $0.82a$ the modal pattern at the lower band-gap edge is the anti-symmetric (odd) mode while that at the upper band-gap edge is the symmetric (even) mode. When the parameter $b$ changes from $0.82a$ to $0.18a$ the modal patterns at the lower and upper band-gap edges interchange. This phenomenon is referred to as the band inversion. Due to the mirror symmetry of the unit-cell, the Zak phase (i.e., the topological invariant) of each isolated band can be determined by the eigenmode symmetry \cite{xiao2015geometric}. For the proposed PnC beam, the Zak phase of the first flexural band is $\pi$ (0) when $b$ is $0.82a$ ($0.18a$). As a result, the topological characteristics of the first flexural mode band-gap are distinct for $b = 0.82a$ and $b = 0.18a$. It is well known that the interface between two PnCs with different topological phases supports the topologically protected interface state. 

The variation of the band-gap edges for the symmetrically flexural mode is shown in Fig. \ref{Figure1}(e). As $b$ diverges from $0.5a$ the lower band-gap edge decreases first until it reaches the minimum, and then increases slightly. Meanwhile, the upper band-gap edge shifts to higher frequencies monotonously. Consequently, the normalized band-gap width increases gradually as $b$ diverges from $0.5a$, as shown in Fig. \ref{Figure1}(f). Furthermore, to reveal the attenuation behavior of the flexural mode, the value of the imaginary part of the wave number Im($k$) is also included in Fig. \ref{Figure1}(e). The wave amplitude attenuates due to the band-gap effect, which can be expressed as $A(x) = \gamma e^{-\mathrm{Im} (k) x},$ where $A$ is the amplitude of the displacement field, $x$ the coordinate along the PnC beam, and $\gamma$ the fitting parameter \cite{romero2010evidences}. It is observed that as the band-gap width increases, the imaginary part of the wave number increases as well. Moreover, the largest value of Im($k$) usually appears near the center of the band-gap.

A PnC ribbon is utilized to study the topologically protected interface state, see Fig. \ref{Figure2}(a). The PnC with $b = 0.82a$ (i.e., crystal 2) is sandwiched by crystal 1 ($b = 0.18a$) on its left and right sides and hence two interfaces are constructed. To calculate the band structure we consider this ribbon structure as a super-cell and then apply the Bloch's boundary conditions on the leftmost and rightmost boundaries. The total length of the super-cell is $24a$ (containing 24 unit-cells). Notably, as we focus our discussion on the symmetrically flexural mode, for the sake of simplicity only half of the structure is employed and the symmetric boundary condition is applied on the mirror plane (i.e., the $y = 0$ plane). Therefore, only the symmetric modes will be calculated in the band structure.

The band structure of the PnC ribbon is shown in Fig. \ref{Figure2}(b). Two degenerate bands corresponding to the topologically protected interface state emerge within the band-gap for the symmetrically flexural mode. Notice that these bands are nearly flat, which means that the interface state is highly localized. From the modal shape (Fig. \ref{Figure2}(c)) it is seen that the flexural waves are confined near the interfaces and decay rapidly away from the interfaces, which confirms the existence of the topological interface state.

\section{Experimental validation}
\label{sec3}

\begin{figure*}[t]
	\centering
	\includegraphics{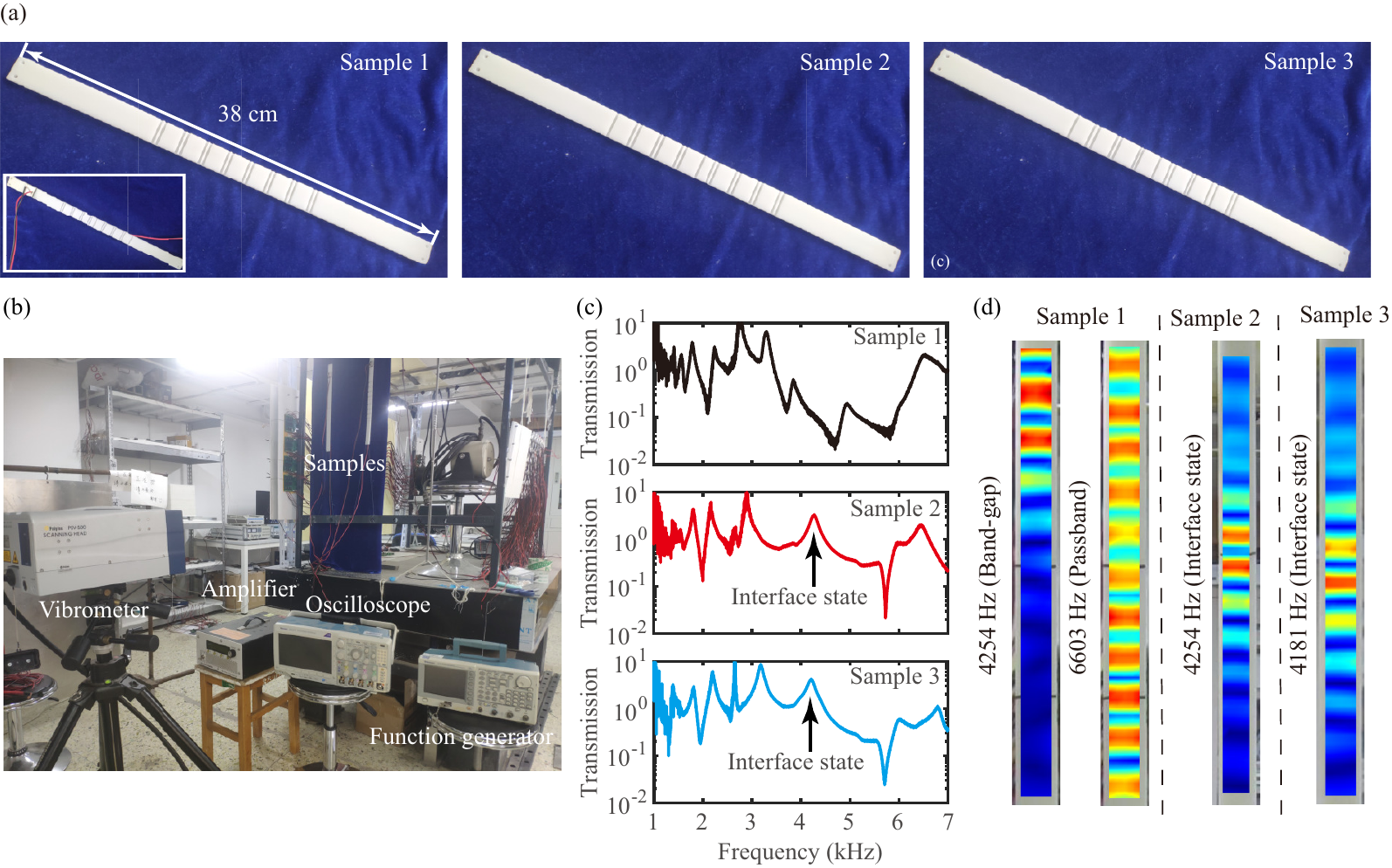}
	\caption{\label{Figure3}(a) Photos of the 3D-printed PnC beams, where the inset shows the piezoelectric patch attached on the sample for the excitation of the flexural waves. (b) Experimental setup for characterizing the PnC beam energy harvester. (c) Measured transmission spectra of the fabricated PnC beams. (d) Measured displacement fields ($|u_z|$) of the fabricated PnC beams at selected frequencies.}
\end{figure*}

To experimentally investigate the flexural wave characteristics of the proposed PnC beams, we fabricate the samples by the 3D-printing technique, see Fig. \ref{Figure3}(a). Herein, each PnC beam contains 8 unit-cells. All of the PnC unit-cells in sample 1 are identical, where the geometrical parameter $b$ is chosen as $0.18a$. For the second PnC beam (sample 2), an interface is constructed between two PnCs ($b = 0.18a$ and $0.82a$, respectively) with distinct topological phases. Thus, it is expected that the topological interface state will appear near the interface. In order to confirm the robustness of the topological interface state against disorders, we introduce the disorders by changing the positions of all the grooves of sample 2, see sample 3 (the right panel of Fig. \ref{Figure3}(a)). It is pointed out that the grooves are randomly shifted by $\delta$, where $\delta$ ranges from $-0.03a$ to $0.03a$ by an increment of $0.01a$.

The photograph of the experimental setup is displayed in Fig. \ref{Figure3}(b). Piezoelectric patches are attached to the PnC beams as the excitation source, see the inset of Fig. \ref{Figure3}(a). The piezoelectric patches are controlled by a function generator (Tektronix, AFG3102) and a power amplifier (T\&C Power Conversion, AG Series Amplifier). Besides, the out-of-plane displacement fields on the back surfaces of the PnC beams are measured and visualized by a laser Doppler vibrometer (Polytec, PSV-500).

The chirp signals over the frequency range of $1 \sim 7$ kHz are employed to test the frequency responses of the fabricated PnC beams. The measured transmission spectra of the three samples are shown in Fig. \ref{Figure3}(c). For sample 1, one can observe that the transmission coefficient significantly drops between 3.2 kHz and 6.4 kHz due to the band-gap effect. These results are consistent with the band structure prediction (Fig. \ref{Figure1}(c)), which confirms the existence of the band-gap. On the other hand, the topological interface state can be identified for both the perfect and perturbed PnC beams with the interfaces (samples 2 and 3), see the transmission peaks inside the band-gap. Therefore, despite the structural perturbation, the interface state is preserved because of the topological protection. However, as the frequency of the interface state is dependent on the actual beam structure, this frequency shifts from 4254 Hz (sample 2) to 4181 Hz (sample 3) by the structural perturbation. Figure \ref{Figure3}(d) plots the measured out-of-plane displacement fields of the PnC beams. Obviously, the flexural wave at the frequency within the band-gap is prohibited. In contrast, the flexural wave within the passband can pass through the PnC beam. Additionally, for either the perfect PnC beam (sample 2) or the perturbed one (sample 3), the modal shapes of the topological interface state show that the flexural wave is confined near the interface and decays rapidly away from the interface. The movies of the scanned displacement fields of the above-mentioned four modes are also provided, see the supplementary material.

\begin{figure*}[t]
	\centering
	\includegraphics{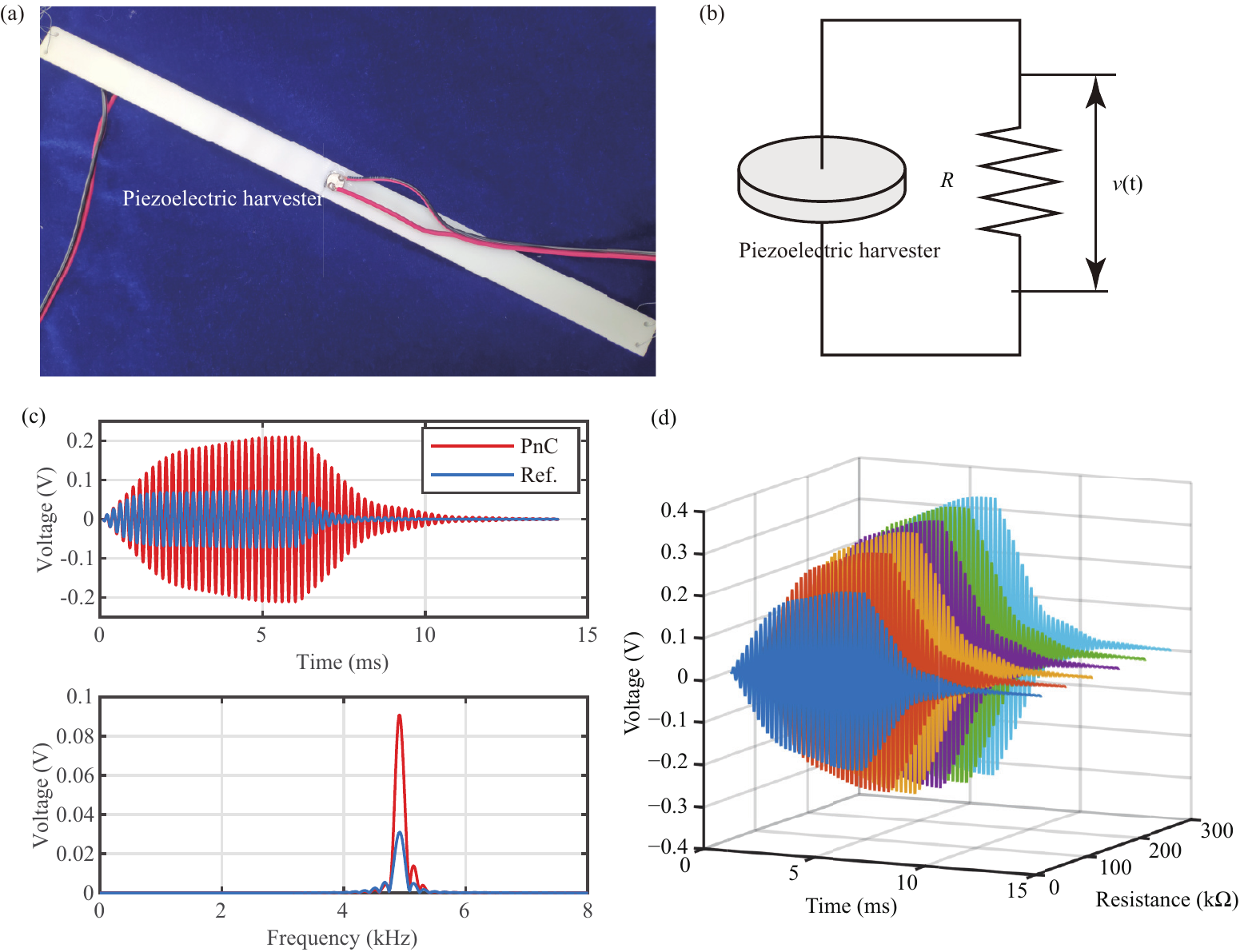}
	\caption{\label{Figure4}(a) Attachment of the piezoelectric harvester at the interface of the PnC beam. (b) Schematic of the electrical circuit of the energy harvesting system. (c) Measured voltage outputs in the time and frequency domains. (d) Measured voltage outputs in the time domain for different electrical load resistances.}
\end{figure*}

Next, we will design a piezoelectric energy harvester based on the topological interface state. To this end, a piezoelectric disc is attached at the interface of the PnC beam possessing the interface state, as shown in Fig. \ref{Figure4}(a). The piezoelectric disc acts as a harvester that converts the elastic wave energy to the electricity. It is worth noting that after the piezoelectric disc is attached, the interface state shifts to a higher frequency owing to the increase of the effective flexural stiffness of the beam. Figure \ref{Figure4}(b) illustrates the electrical circuit of the energy harvester system, where an electrical resistance is connected to the harvester. The output voltage over the electrical resistance is measured by using an oscilloscope (Tektronix, MDO3014). To validate the PnC energy harvester, the 30-cycle sinusoidal signal at the frequency corresponding to the interface state is used.

The measured voltage from the piezoelectric harvester in the perfect PnC beam (sample 2) as a function of time is shown in the top panel of Fig. \ref{Figure4}(c). Moreover, a bare beam (i.e., without grooves) is 3D-printed and measured as a reference, and the output from this beam (blue solid line) is plotted as well. It is apparent that the output voltage from the PnC beam energy harvester is higher than that from the reference beam harvester. This is because the mechanical displacement amplitude near the interface of the PnC beam is amplified by the topological interface state. The corresponding voltages in the frequency domain are also plotted in Fig. \ref{Figure4}(c) (bottom panel). One can see that the harvested voltage from the PnC beam is about three times as high as that from the reference beam. 

It is noteworthy that the electrical load resistance affects the voltage and power outputs of the PnC beam harvesters. The measured voltage outpurts from the perfect PnC beam harvester (sample 2) for different load resistances are shown in Fig. \ref{Figure4}(d). As the load resistance increases, the amplitude of the output voltage increases as well.

\begin{figure*}[t]
	\centering
	\includegraphics{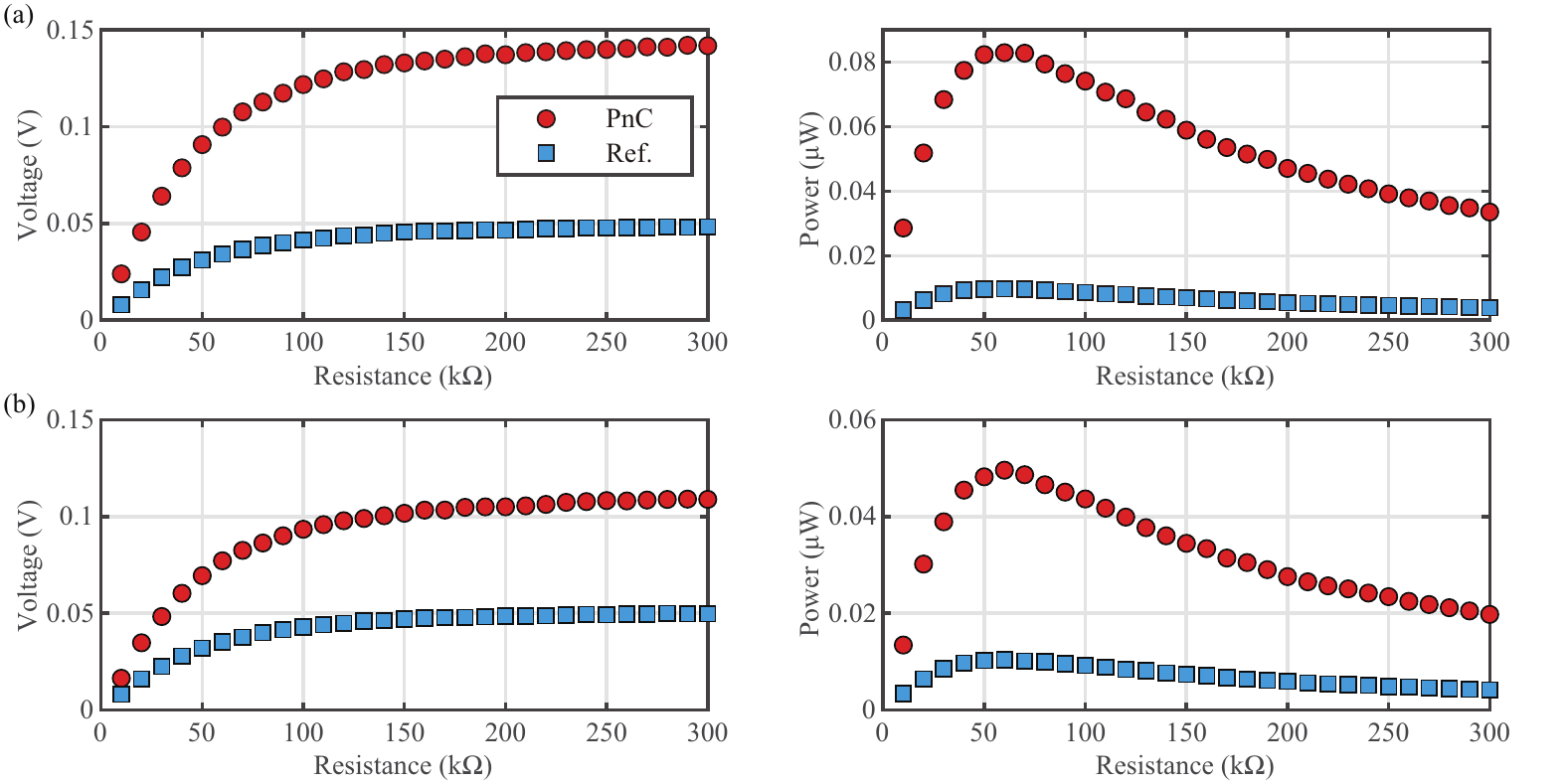}
	\caption{\label{Figure5}Voltage and power outputs obtained from the perfect (a) and perturbed (b) PnC beams as a function of the electrical load resistance, where the excitation frequencies correspond to those of the interface state of the perfect and perturbed PnC beams, respectively. The electrical outputs obtained from the reference beam are also plotted.}
\end{figure*}

The electrical outputs generated by the PnC and reference harvesters as functions of the electrical load resistance are then presented in Fig. \ref{Figure5}, see the red circles and the blue squares, respectively. For the perfect PnC beam (sample 2), it is seen that as the load resistance increases, the measured voltage rises monotonously but its increasing rate declines gradually. Obviously, compared with the reference beam the output voltage is amplified by the PnC beam. The power generated by the piezoelectric harvesters can be calculated according to the Ohm's law. For both the PnC and reference beams, as the load resistance increases, the power outputs grow until they reach their maximum values, and thereafter they begin to decline, as shown in the right panel of Fig. \ref{Figure5}(a). The maximum power generated by the PnC beam harvester is 0.083 $\mu \rm{W}$. Herein, the amplification ratio is defined as the ratio of the power outputs obtained from the PnC beam and the reference beam. The measured amplification ratio of the perfect PnC beam harvester reaches 8.5.

Similar phenomena can be observed in the case of the perturbed PnC beam harvester (sample 3), as shown in Fig. \ref{Figure5}(b). Although the random perturbations are introduced into the PnC beam, the topologically interface state still exists. Consequently, the output electric signal from the perturbed PnC beam is larger than that from the reference beam. However, it should be pointed out that the energy harvesting performance of the perturbed PnC beam degrades due to the introduction of the structural disorders. The amplification ratio of the perturbed beam harvester drops to 4.8. In the case of the perturbed PnC beam, the piezoelectric disc is attached at the same position as for the perfect PnC beam. It is found that, for the perfect and perturbed PnC beams, the positions of the maximum displacement amplitude of the interface state are slightly different (Fig. \ref{Figure3}(d)). This results in the performance degradation of the piezoelectric harvester based on the perturbed PnC beam.

\section{Conclusions}
\label{sec4}

In this paper, we design the PnC beam with the topologically protected interface state for harvesting the mechanical energy of the flexural waves. The topological interface state is generated by the combination of the 1D PnCs with topologically trivial and non-trivial phases. We fabricate the PnC samples by 3D printing and then demonstrate numerically and experimentally the existence of the topological interface state. Based on the topological interface state, the PnC beam for the energy harvesting application is validated. For this purpose, a piezoelectric disc is attached at the interface of two topologically different PnCs as the energy harvester. Compared to the reference beam harvester, the output power is amplified by the PnC harvester at the frequency of the interface state. Under the optimal electrical load resistance, the measured amplification ratio of the PnC beam harvester is 8.5. Moreover, although the random disorders are introduced into the perfect PnC beam, the enhancement of the energy harvesting can also be achieved by using the topological interface state. 

\begin{acknowledgments}
The work is supported by the German Research Foundation (DFG, ZH 15/27-1), and the Joint Sino-German Research Project (Grant No. GZ 1355). Y.-S. Wang is also grateful to the support by innovative research group of NSFC (12021002) and Major Program of National Science Foundation of China (11991031).
\end{acknowledgments}


\end{document}